\documentstyle[11pt,amssymb]{article}

\makeatletter
\@addtoreset{equation}{section}
\makeatother

\def\dalemb#1#2{{\vbox{\hrule height .#2pt
        \hbox{\vrule width.#2pt height#1pt \kern#1pt
                \vrule width.#2pt}
        \hrule height.#2pt}}}

\def\cF{{\cal F}}
\def\cA{{\cal A}}

\def\0{{\sst{(0)}}}
\def\1{{\sst{(1)}}}
\def\2{{\sst{(2)}}}
\def\3{{\sst{(3)}}}
\def\4{{\sst{(4)}}}
\def\5{{\sst{(5)}}}
\def\6{{\sst{(6)}}}
\def\7{{\sst{(7)}}}
\def\8{{\sst{(8)}}}

\def\tH{\widetilde H}

\def\Z{\rlap{\sf Z}\mkern3mu{\sf Z}}
\def\R{\rlap{\rm I}\mkern3mu{\rm R}}

\def\wtd{\widetilde}

\def\Qw{{Q_{\rm wave}}}
\def\Qnut{{Q_{\sst{\rm NUT}}}}
\def\mun{{\mu_{\sst{\rm NUT}}}}
\def\muw{{\mu_{\rm wave}}}
\let\a=\alpha

 \let\ci=\cite 
   
\def\nn{\nonumber} \def\bd{\begin{document}} \def\ed{\end{document}}
\def\ds{\documentstyle} \let\fr=\frac \let\bl=\bigl \let\br=\bigr
\let\Br=\Bigr \let\Bl=\Bigl 
\let\bm=\bibitem
\let\na=\nabla
\let\pa=\partial \let\ov=\overline 
\newcommand{\be}{\begin{equation}} 
\newcommand{\ee}{\end{equation}} 
\def\ba{\begin{array}}
\def\ea{\end{array}}
\def\ft#1#2{{\textstyle{{\scriptstyle #1}\over {\scriptstyle #2}}}}
\def\fft#1#2{{#1 \over #2}}
\def\del{\partial}
\def\sst#1{{\scriptscriptstyle #1}}
\def\oneone{\rlap 1\mkern4mu{\rm l}}
\def\ie{{\it i.e.\ }}
\def\via{{\it via}}
\def\semi{{\ltimes}}
\def\str{{\rm str}}
\def\jm{{\rm j}}
\def\im{{\rm i}}

\newcommand{\ho}[1]{$\, ^{#1}$}
\newcommand{\hoch}[1]{$\, ^{#1}$}
\newcommand{\bea}{\begin{eqnarray}} 
\newcommand{\eea}{\end{eqnarray}} 
\newcommand{\ra}{\rightarrow}
\newcommand{\lra}{\longrightarrow}
\newcommand{\Lra}{\Leftrightarrow}
\newcommand{\ap}{\alpha^\prime}
\newcommand{\bp}{\tilde \beta^\prime}
\newcommand{\tr}{{\rm tr} }
\newcommand{\Tr}{{\rm Tr} } 
\newcommand{\NP}{Nucl. Phys. }
\newcommand{\tamphys}{\it Center for Theoretical Physics\\
Texas A\&M University, College Station, Texas 77843}
\newcommand{\ens}{\it Laboratoire de Physique Th\'eorique de l'\'Ecole
Normale Sup\'erieure\hoch{2,3}\\
24 Rue Lhomond - 75231 Paris CEDEX 05}
\newcommand{\upenn}{\it Department of Physics and Astronomy\\
University of Pennsylvania, Philadelphia, Pennsylvania 19104}

\newcommand{\auth}{M. Cveti\v{c}\hoch{\dagger1},
H. L\"u\hoch{\dagger1} and C.N. Pope\hoch{\ddagger2}}

\begin{document}
\begin{flushright}
\hfill{CTP TAMU-44/98}\\
\hfill{UPR-821-T}\\
\hfill{hep-th/9811107}\\
\hfill{November 1998}\\
\end{flushright}


\begin{center}
{ \large {\bf Decoupling Limit, Lens Spaces and Taub-NUT:\\
D=4 Black Hole Microscopics from D=5 Black Holes}}

\vspace{10pt}
\auth

\vspace{10pt}

{\hoch{\dagger}\upenn}

\vspace{10pt}
{\hoch{\ddagger}\tamphys}

\vspace{40pt}

\underline{ABSTRACT}
\end{center}

We study the space-times of non-extremal intersecting $p$-brane
configurations in M-theory, where one of the components in the
intersection is a ``NUT,'' \ie a configuration of the Taub-NUT type.
Such a Taub-NUT configuration corresponds, upon compactification to
$D=4$, to a Gross-Perry-Sorkin (GPS) monopole.  We show that in the
decoupling limit of the CFT/AdS correspondence, the 4-dimensional
transverse space of the NUT configuration in $D=5$ is foliated by
surfaces that are cyclic lens spaces $S^3/Z_N$, where $N$ is the
quantised monopole charge. By contrast, in $D=4$ the 3-dimensional
transverse space of the GPS monopole is foliated by 2-spheres.  This
observation provides a straightforward interpretation of the
microscopics of a $D=4$ string-theory black hole, with a GPS monopole
as one of its constituents, in terms of the corresponding $D=5$ black
hole with no monopole.  Using the fact that the near-horizon region of
the NUT solution is a lens space, we show that if the effect of the
Kaluza-Klein massive modes is neglected, $p$-brane configurations can
be obtained from flat space-time by means of a sequence of dimensional
reductions and oxidations, and U-duality transformations.

{\vfill\leftline{}\vfill
\footnoterule
{\footnotesize \hoch{1} Research supported in part by DOE grant 
DE-FG02-95ER40893 \vskip -12pt} \vskip 10pt
{\footnotesize \hoch{\phantom{1}} and University of Pennsylvania 
Research Foundation.\vskip -12pt} \vskip 14pt
{\footnotesize  \hoch{2} Research supported in part by DOE 
Grant DE-FG03-95ER40917.\vskip  -12pt}}

\pagebreak
\setcounter{page}{1}

\section{Introduction\label{sec:intro}}

      Advances in the quantitative treatment of black-hole
microscopics \cite{vs} represent an important ``spin-off'' of 
M-theory unification, facilitated by developments in the quantum
treatment of non-pertubative objects in string theory, such as
$D$-branes \cite{Polchinski}.

      More recently, aspects of the black-hole microscopics have found
an elegant reinterpretation within the framework of the CFT/AdS
correspondence~\cite{Maldacena}.  Namely, the microscopic
interpretation of black hole entropy can be made quantitative in terms
of the boundary conformal field theory, as determined by the anti-de
Sitter space-time, \ie the (asymptotic) geometry in the decoupling
limit of certain black holes. In particular, the near-extremal black
holes in $D=5$ \cite{Strominger,cvla} (or $D=4$) \cite{bala,cvlaII}
have a six-dimensional (or five-dimensional) embedding \cite{hyun,SS}
as black strings, with or without rotation, whose geometry in the
decoupling limit is BTZ$\times S^{3}$ (or BTZ$\times S^{2}$), where
BTZ denotes the Ba\~nados-Teitelboim-Zanelli three-dimensional black
hole space-time, which is locally AdS$_3$ \cite{btz}.  Thus its
quantum states are determined asymptotically by a two-dimensional
conformal field theory at the asymptotic boundary \ci{adsc}.  The
counting of states in this CFT is then used
\cite{Strominger,dublinbtz} to reproduce the Bekenstein-Hawking
entropy.  (The analysis of black holes in $D\ge 6$ involves a
correspondence to CFT's in $D>2$, and there, due to renormalisation
effects \cite{guklts} in strongly coupled CFT's, only a qualitative
modelling of the microscopic black hole entropy is possible
\cite{klts,cvlupo}.)  Further study of the microscopic black hole
spectrum both on the gravity AdS$_3$ side \cite{Sezgin}, as well as on
the CFT$_2$ side \cite{mast,deBoer,Larsen}, has been pursued.
 
     This paper addresses a number of related topics.  In section 2,
we study the near-horizon geometry that is relevant in the decoupling
limit for extremal and non-extremal $p$-branes in $M$-theory, in the
case where one of the ingredients in the intersection is a NUT, \ie a
configuration of the Taub-NUT type.  Such intersections become
four-dimensional black holes upon dimensional reduction on $T^7$, with
the Taub-NUT component corresponding to a magnetic charge carried by
one of the Kaluza-Klein vectors.\footnote{In general, the
higher-dimensional configuration with the topological twist in the
fibre coordinate associated with the Kaluza-Klein vector carrying the
magnetic charge will be described, for brevity, as a NUT.}  If the
Kaluza-Klein vector is the one coming from the reduction step from
$D=5$ to $D=4$, this corresponds to the situation arising in the
Gross-Perry-Sorkin (GPS) monopole (\ie a $D=4$ black hole where the
Kaluza-Klein vector from $D=5$ carries a magnetic charge
\cite{grpeso}).  If, on the other hand, we consider eleven-dimensional
configurations with the same set of intersecting components, but with
no NUT, they will now already be interpretable as black holes (as
opposed to Taub-NUT configurations) in $D=5$, with one less charge
than that in $D=4$.  (This will be discussed further in section
2.)\footnote{Note that when one speaks of a configuration of $N$
intersecting objects, this is not the same thing as a configuration of
$(N+1)$ intersecting objects in which the $(N+1)$'th charge is set to
zero.  The reason for this is that the individual ingredients in an
intersection are distributed uniformly over the world-volumes of the
other ingredients.  For example, in the case where one sets
the charges of all but one of the ingredients in an intersection to
zero, the remaining configuration will not be a single isotropic
object, but instead it will be smeared uniformly over all the spatial
world-volume directions of the (now-vanished) other ingredients.  Thus
when we speak of an eleven-dimensional configuration ``but with no
NUT,'' we mean the configuration that one would have considered if the
NUT were never introduced, as opposed to the the configuration that
would result from setting the NUT charge to zero in the original
intersection.}  The crucial observation is that in the decoupling
limit, the foliating 3-surfaces in the transverse space of the NUT
solution have the geometry of a cyclic lens space, with the topology
$S^3/Z_N$, where $N$ is the quantised NUT charge.  This observation
allows us to interpret, {\it via} the AdS/CFT correspondence, the
microscopics of such $D=4$ black holes in terms of $D=5$ black holes
with one less charge.
  
      In section 3, we show that by performing a sequence of
dimensional reductions and oxidations on the $U(1)$ fibres of the
foliating 3-spheres of a flat 4-dimensional transverse space,
supplemented by appropriate U-duality transformations, we can generate
$p$-brane configurations in M-theory from the flat space.  This
provides a new way of generating BPS solutions, by starting from a
flat space-time and just using symmetries of the theory as a
solution-generating procedure.

\section{Decoupling limit and microscopics of $D=4$ black-hole states
 with GPS-monopole\label{sec:micro}}

In four-dimensional maximal supergravity, which is the effective
low-energy limit of $M$-theory compactified on the 7-torus, black-hole
solutions form multiplets under the $E_{7(+7)}$ U-duality group.  The
prototype solution is specified by four charges. (The generating
solution of the most general black hole, consistent with the no-hair
theorem, is actually specified by five charges \cite{cvhu}; however
the global space-time features are essentially captured by the
4-charge solution.) In the ``diagonal'' case where each of the charges
is associated with a specific harmonic function (\ie a {\it generating
solution}), the solutions have a simple structure, and are referred to
as four-charge solutions (first specified by two electric and two
magnetic charges of the Neveu-Schwarz-Neveu-Schwarz (NS-NS) sector and
given explicitly in \cite{cvyo}).  The possible field strength
configurations that can give rise to such simple 4-charge solutions
are given by \cite{classp}
\bea
N=4:&& \{ F_{\2 ij}, F_{\2 k\ell}, F_{\2 mn}, *{\cal F}_\2^p
        \}_{105+105}\ ,\quad
       \{F_{\2 ij}, *F_{\2 ik}, {\cal F}_\2^j, *{\cal
         F}_\2^k \}_{210}\ ,\nonumber\\
    && \{ F_{\2 ij}, F_{\2 k\ell}, *F_{\2 ik}, *F_{\2 j\ell}
        \}_{210}\ .\label{d4n4}
\eea
(We are using the notation of \cite{lpsol,cjlp1,classp} here.)  The
subscripts on each bracketed set of field strengths denotes the
multiplicities of the solutions (corresponding to the possible
permutations of index choices on the field strengths).  The Hodge
duals indicate that the associated fields carry electric charges if
the fields without duals carry magnetic charges, and {\it vice versa}.

     When oxidised  back to $D=6$, there are four possible
near-horizon limits that can arise, namely
\be
\hbox{AdS$_3\times S^3$,
AdS$_3\times (S^2\times S^1)$,
(AdS$_2\times S^1)\times S^3$,
(AdS$_2\times S^1) \times (S^2\times S^1)$}\ .
\label{horizons}
\ee
(To be precise, the AdS$_3$ refers only to the local space-time, which
can be globally that of the BTZ black hole, and also the $S^3$ can in
general be squashed and/or factored by a cyclic group, in the manner
described in \cite{dlpads} and in subsequent discussions.)  If we
oxidise these near-horizon solutions further, to $D=10$ or $D=11$,
then the additional dimensions provide further factors of $T^4$ or
$T^5$ respectively.  These near-horizon geometries are related to each
other by Hopf T-duality, which is a T-duality that makes use of
the $U(1)$ isometry of the fibre bundle coordinate over the base space
\cite{dlpads5}.  The U-duality and T-duality transformations that relate
the different topologies in (\ref{horizons}) leave the areas of the
horizons invariant, and they may therefore be called isentropic
mappings \cite{dlpads}.

      The first of the three cases in (\ref{d4n4}) can be viewed in
$D=11$ as the intersection of three M2-branes and one NUT, or else the
magnetic dual of this, namely, three M5-branes and a gravitational 
pp-wave~\cite{cvtsII}.  (The microscopic state counting in the context of
the AdS/CFT correspondence has been given in \cite{bala} for static
black holes, and in \cite{cvlaII} for rotating black holes.) The third
case in (\ref{d4n4}) can be viewed as the intersection of two
M2-branes and two M5-branes~\cite{cvtsII}.

   In this paper, we shall concentrate on the second case in
(\ref{d4n4}).  These configurations can be viewed as intersections of
an M2-brane and an M5-brane, together with a wave and a NUT.  In
particular, we shall consider the case where the indices $j$ and $k$
take the values $\{j,k\}=\{6,7\}$, so that the solution can be oxidised
back to $D=6$ to become~\cite{cvts} a dyonic string
\cite{Rahmfeld,DKL,DLPhet,DLPgauge} with a pp-wave propagating on its
word-sheet, and a NUT planted in its transverse space. If
the index $i$ takes any of the values $i=2$, 3, 4, or 5, the dyonic
string belongs to the R-R sector, If instead the index $i$ takes the
value $i=1$, the dyonic string belongs to the NS-NS sector~\cite{cvts}. 
In this latter case, the discussion is equally applicable to the
heterotic string.

\subsection{Extremal case}

    Although the extremal solution can be obtained by taking the
extremal limit of the non-extremal solution, their respective
geometries are quiet different and we shall discuss them
separately. We begin with the extremal solution.  The extremal
4-charge black hole for the field configuration $\{F_{\2 i6}, {*F}_{\2
i7}, \cF^6_\2, *\cF^7_\2\}$ is a solution of the bosonic Lagrangian
\be
\kappa_{4}^2\, e^{-1}\, {\cal L} =
R-\ft12(\del\vec \phi)^2 -
\ft14 e^{\vec a_{i6}\cdot \vec\phi}\, (F_{\2 i6})^2 -
\ft14 e^{\vec a_{i7}\cdot \vec\phi}\, (F_{\2 i7})^2 -
\ft14 e^{\vec b_6\cdot \vec\phi}\, (\cF_{\2}^6)^2 -
\ft14 e^{\vec b_7\cdot \vec\phi}\, (\cF_{\2}^7)^2\ ,
\ee
where the index $i$ can be any number from 1 to 5.  The dilaton
vectors $\vec c_\a =\{ \vec a_{i6}, \vec a_{i7}, \vec b_6, \vec b_7\}$
can be found in \cite{lpsol,cjlp1}; they satisfy \cite{lpmulti}
\be
\vec c_\a \cdot \vec c_\beta =4 \delta_{\a\beta} -1\ .
\ee
Four-charge black hole solutions were constructed in \cite{cvyo}.  In
this case, the solution is given by
\bea
&&ds_{4}^2 = -(H_e\, H_m\, K\, U)^{-1/2}\, dt^2 +
            (H_e\, H_m\, K\, U)^{1/2}\, 
(d\rho^2 + \rho^2\, d\Omega_2^2)\ ,\nn\\
&&A_{\1 16} = H_e^{-1}\, dt\ ,\quad
F_{\2 17} =Q_m\, \Omega_\2 ,\quad
\cA^6_\1 = K^{-1}\, dt\ ,\quad
\cF^7_{\2} =\Qnut\, \Omega_\2\ ,\nn\\
&&\vec \phi = \ft12 \vec a_{16}\, \log H_e -
\ft12 \vec a_{17}\, \log H_m +
\ft12 \vec b_6 \log K -\ft12 \vec b_7 \log U\ ,
\label{d4bh}
\eea
where the harmonic functions are 
\be
H_e=1 + \fft{Q_e}{\rho}\ ,\qquad
H_m=1 + \fft{Q_m}{\rho}\ ,\qquad
K=1 + \fft{\Qw}{\rho}\ ,\qquad
U= 1 + \fft{\Qnut}{\rho}\ .\label{d4harmonic}
\ee

     The metric of the extremal 4-charge black hole (\ref{d4bh}) in
$D=4$ has a regular  horizon at $\rho=0$, near to which the 
geometry approaches AdS$_2\times S^2$. The entropy of the
solution is 
\bea
S&\equiv&\fft{\hbox{Area}}{4\kappa_{4}^2}\nn\\
&=& \fft{\omega_2}{4\kappa_4^2}\, \sqrt{Q_e\, Q_m\, \Qw\, \Qnut}\ .
\label{d4ent}
\eea
Here $\omega_2=4\pi$ is the volume of the unit two-sphere.  (In this
paper we use $\omega_n$ to denote the volume of the unit $n$-sphere,
and $\Omega_{(n)}$ to denote its volume form.  Thus $\omega_n =
\int\Omega_{(n)}$.  We denote the metric of the unit $n$-sphere by
$d\Omega_n^2$.)

\subsubsection{$D=4$ and $D=5$ black holes}

    First, let us consider the oxidation of the 4-charge black-hole
solution (\ref{d4bh}) to $D=5$.  Since the internal coordinate
associated with this step is $z_7$, which we shall denote by $y$, 
it follows that the charges $Q_e$,
$Q_m$ and $\Qw$ will now be supported by the field strengths $F_{\2i6}$,
${*F_{\3i}}$ and $\cF^6_\2$ in $D=5$.  However the charge $\Qnut$,
being associated with the Kaluza-Klein vector $\cA_\1^7$ of the $D=5$
to $D=4$ reduction step, becomes instead a topological ``NUT charge''
in $D=5$.  The metric in (\ref{d4bh}) oxidises to become
\be
ds_5^2 = -(H_e\, H_m\, K)^{-2/3}\, dt^2 + (H_e\, H_m\, K)^{1/3})\,
\Big[ U\, (d\rho^2 + \rho^2\, d\Omega_2^2) + U^{-1}\, (dy + \Qnut\,
B)^2 \Big]\label{d5met}
\ee
in five dimensions, where $B$ is a 1-form on the unit 2-sphere, such
that $dB=\Omega_\2$.

    This is an appropriate juncture at which to comment further on the
observation made in footnote 2.  If we set the NUT charge $\Qnut$ in
(\ref{d5met}) to zero, then we get
\be
ds_5^2 = -(H_e\, H_m\, K)^{-2/3}\, dt^2 + (H_e\, H_m\, K)^{1/3})\,
\Big[ d\rho^2 + \rho^2\, d\Omega_2^2 + dy^2  \Big]\ .\label{d5metwon}
\ee
Although this ostensibly looks like a standard 3-charge black hole in
$D=5$ it is actually a {\it line} of 3-charge black holes, since the
remaining harmonic functions $H_e$, $H_m$ and $K$ depend only on
$\rho$, rather than on the entire radial coordinate $R=\sqrt{\rho^2 +
y^2}$ in the 4-dimensional transverse space: A standard isotropic
3-charge black hole would have harmonic functions of the form $1+
4Q/R^2$, rather than $1+Q/\rho$.  In fact the line of black holes
described by the metric (\ref{d5metwon}) is precisely what would
result from performing a normal vertical oxidation of an isotropic
3-charge black hole in $D=4$.  As we shall now show, if we instead
perform the oxidation when $\Qnut$ is an additional non-vanishing
fourth charge in a $D=4$ black hole, with $y$ the fibre coordinate of
the $U(1)$ bundle associated with the NUT charge, we instead arrive at
a configuration that {\it does} correspond to an isotropic 3-charge
black hole in $D=5$.

     If we go to a region near the horizon, defined by the
requirement that $\rho << \Qnut$ so that we can drop the ``1'' in the
harmonic function $U=1 + \Qnut/\rho$, and if we also define a new
radial coordinate $r$ by $\rho= r^2/4$, we find that the
five-dimensional metric (\ref{d5met}) can be approximated as
\be
ds_5^2 = -(H_e\, H_m\, K)^{-2/3}\, dt^2 + \Qnut\, (H_e\, H_m\, K)^{1/3})\,
\Big[dr^2 + \ft14 r^2\, d\Omega_2^2 + \ft14 r^2 (\fft{dy}{\Qnut} +
B)^2 \Big]\ ,\label{d5met2}
\ee
The metric
\be
d\Omega_3^2(\Qnut)\equiv  \ft14 d\Omega_2^2 + 
\ft14 (\fft{dy}{\Qnut} + B)^2\label{lensmetric}
\ee
is locally the standard metric on the unit 3-sphere.  In fact it would
be precisely the unit metric on $S^3$ if $y$ had the period $4\pi\,
\Qnut$.  If instead $y$ has the period $4\pi$, then the metric
describes the cyclic lens space $S^3/Z_\Qnut$, where the fibre
coordinate of the $U(1)$ bundle over the $S^2$ base is
identified\footnote{Note that the identification has no fixed points,
and thus $S^3/Z_\Qnut$ is a smooth manifold, and not an
orbifold. However, if one considers the flat-space metric
$$ ds^2=dr^2+ \ft14 r^2\, d\Omega_2^2 + \ft14 r^2 (\fft{dy}{\Qnut}
+B)^2\ ,
$$
and takes $y$ to have the period $4\pi$, then there is a
fixed point at the origin when the integer $\Qnut$ is not equal to
unity,  where the foliating lens spaces shrink down to a point.  This
gives rise to an orbifold singularity in the manifold $R^4/Z_\Qnut$.}
after a translation by a fraction $1/\Qnut$ of its total length in
$S^3$.  (The charges are normalised here so that the Dirac
quantisation condition requires that they be integers.)  In fact we
are obliged to take $y$ to have the period $4\pi$, rather than
$4\pi\, \Qnut$, since we require that four-dimensional black
holes, for {\it any} integer value of the charge $\Qnut$, should be
oxidisable to regular geometries in 
$D=5$.  If we were instead to took the period of $y$ to be
$4\pi \, \Qnut$ for some value of $\Qnut\ne1$, then effectively this
value of $\Qnut$ would itself define the minimum allowed ``unit'' of
charge, and no smaller values would be permitted.  This is because
one cannot have a regular geometry whose $U(1)$ fibres exceed the length
that corresponds to the case of $S^3$ itself.  Only integer fractions
of the length of the fibres for $S^3$ give regular geometries.

    Rewriting the metric (\ref{d5met2}) in terms of
$d\Omega_3^2(\Qnut)$, we obtain
\be
ds_5^2 =  -(H_e\, H_m\, K)^{-2/3}\, dt^2 + \Qnut\, 
(H_e\, H_m\, K)^{1/3})\,
(dr^2 + r^2\, d\Omega_3^2(\Qnut)) \ .\label{d5met3}
\ee
As a result of the coordinate transformation $\rho=r^2/4$, the functions
$H_e$, $H_m$ and $K$ given in (\ref{d4harmonic}), which were harmonic in
the original 3-dimensional transverse space, are now given by
\be
H_e =1 + \fft{4Q_e}{r^2}\ ,\qquad 
H_m =1 + \fft{4Q_m}{r^2}\ ,\qquad 
K =1 + \fft{4\Qw}{r^2}\ .\label{harm5}
\ee
These are harmonic with respect to the 4-dimensional transverse space.
Note that in this ``Hopf'' oxidation on the $U(1)$ fibres, unlike a
standard vertical oxidation in the transverse space, the harmonic
functions are still isotropic in the higher dimension.  (By contrast,
in the usual vertical oxidation the harmonic functions would describe
smeared lines of charge in the higher dimension.)  The
five-dimensional solution can be recognised as having the structure of
the isotropic 3-charge black hole, at least if $y$ is identified with
period $4\pi\, \Qnut$.

\subsubsection{$D=6$ dyonic string with pp-wave and Taub-NUT charge}

        In order to give a microscopic interpretation for the above
semi-classical Hawking entropy, we shall first oxidise the solution
back to $D=6$, where it describes \cite{cvts} a dyonic string,
together with a pp-wave and a NUT.  As mentioned previously, the
dyonic string will be supported by charges in the NS-NS sector if the
index $i$ takes the value $i=1$, and otherwise it will be supported by
charges in the R-R sector.  The associated six-dimensional Lagrangian
is of the form $\kappa_6^2\, e^{-1}\, {\cal L} = R - \ft12
(\del\phi)^2 -\ft14 e^{-\phi}\, (F_\3)^2$.  The six-dimensional
metric, which is independent of the choice of $i$ (\ie it is the same
whether the fields are in the R-R or the NS-NS sector), is obtained by
performing a further step of dimensional oxidation of (\ref{d5met}),
leading to \cite{cvts}:
\bea
ds_{6}^2 &=& (H_e\, H_m)^{-1/2}\, \Big(-K^{-1}\, dt^2 + 
K (dx + (K^{-1}-1) dt)^2 \Big) \nn\\
&+& (H_e\, H_m)^{1/2}\, \Big(U(d\rho^2 + \rho^2\, d\Omega_2^2)
+ U^{-1}(dy + \Qnut\, B)^2\Big)\ ,\label{d6metric1}
\eea
where again $dB=\Omega_\2$ is  the volume form of the unit 2-sphere.
Here, $x=z_6$ is the compactification coordinate of the $S^1$
reduction step from $D=6$ to $D=5$.

As we shall see presently, the area of the horizon (at $r=0$) for the
above metric implies that the entropy of this $D=6$ boosted
dyonic string with NUT charge is:
\bea
S&\equiv& \fft{\hbox{Area}}{4\kappa_{6}^2}\nn\\
&=& \fft{1}{4\kappa_6^2}\, 
((Q_e\, Q_m)^{1/4}\, \Qw^{1/2}\, L_x)\,\, 
(2\Qnut^{1/2}\, (Q_e\, Q_m)^{1/4})^3\,\, \omega_3/\Qnut\nn\\
&=& \fft{2\omega_3\, L_x}{\kappa_6^2}\,
\sqrt{Q_e\, Q_m\, \Qw\, \Qnut}\ . \label{d6ent}
\eea
Note that we have taken the period of the internal coordinate $x$ to
be $L_x$.  The period of the internal coordinate $y$ for the reduction
step from $D=5$ to $D=4$ is $4\pi$, and so
the volume of the internal 2-torus is $4\pi\, L_x$.  It then follows
that the gravitational constants $\kappa_4$ and $\kappa_6$ in $D=4$
and $D=6$ satisfy the following relationship:
\be
\kappa_6^2 = 4\pi\, L_x\, \kappa_4^2\ .\label{d6d4kappa}
\ee
Taking into account the fact that $\omega_3 = \pi\, \omega_2/2$, it
follows that the entropy (\ref{d6ent}) of the $D=6$ string (with the
Taub-NUT-charge) is the same as (\ref{d4ent}), that of the $D=4$ black
hole.  This result is of course a natural consequence of the fact that
entropy is preserved under dimensional reduction.

\subsubsection{Near-horizon region and counting of microstates}

     We now turn to the near-horizon region $\rho\to 0$, which in turn
corresponds to the gravity decoupling limit in the AdS/CFT
correspondence:
\be 
\rho\ll(Q_e,\ Q_m, \ Q_{\rm NUT})\ .\label{dec}
\ee
Note that this  limit does {\it not} impose any restriction on the value
of $Q_{\rm wave}$, relative to $\rho$. The metric  takes the following
form:
\bea
ds_6^2 &=& \fft{r^2}{4\sqrt{Q_e\, Q_m}}\, \Big(
-\fft{r^2}{r^2 + 4\Qw}\, dt^2 +
(1 + \fft{4\Qw}{r^2})\, (dx - \fft{4\Qw}{r^2 - 4\Qw}\, dt)^2\Big)\nn\\
&& + 4\sqrt{Q_e\, Q_m}\, \Qnut\, \fft{dr^2}{r^2} +
4\sqrt{Q_e\, Q_m}\, \Qnut\, d\Omega_3^2(\Qnut)\ ,
\label{d6horizon}
\eea
where we have, as previously, made the coordinate transformation
$\rho=r^2/4$, and $d\Omega_3^2(\Qnut)$ denotes the metric
(\ref{lensmetric}) on the unit cyclic lens space $S^3/Z_\Qnut$, where
the coordinate $y$ has the period $4\pi$. Note that the volume of the
unit-radius lens space is $\omega_3/\Qnut$, where $\omega_3$ is the
volume of the unit-radius three-sphere.  The fact that the space
associated with the line element $d\Omega_3^2(\Qnut)$ is a lens space
with the topology $S^3/Z_N$, where $N$ is the quantised value of the
charge $Q_{\rm NUT}$, has important implications for the microscopic
interpretation. The metric (\ref{d6horizon}) describes a direct
product of two three-dimensional spaces, namely the extremal BTZ black
hole \cite{btz} and the lens space $S^3/Z_\Qnut$.  To see this, we
first make the coordinate rescaling $r\rightarrow r/\sqrt{\Qnut}$, and
then we dimensionally reduce the six-dimensional metric
(\ref{d6horizon}) on the lens space.  Specifically, we take the lens
space metric for the compactification to be scaled to $ds^2
=4\sqrt{Q_e\, Q_m}\, \Qnut\, d\Omega_3^2(\Qnut)$.  With this choice
for the internal metric, the (constant) breathing-mode scalar takes
the value 0, and hence there is no conformal rescaling of the
space-time metric.  (See \cite{bdlps} for a detailed discussion of
Kaluza-Klein reduction on spheres and other spaces.)  The resulting
three-dimensional space-time metric is then given by
\be
ds_3^2=-\fft{r^4}{\ell^2\, R^2}\, dt^2 +
R^2\, (\fft{dx}{\ell} - \fft{2\kappa_3^2\, J}{R^2}\, dt)^2 + \ell^2\, 
\fft{dr^2}{r^2}\ ,\label{btzextremal}
\ee
where
\bea
&&J=M\, \ell\ ,\qquad
\ell=2(Q_e\, Q_m)^{1/4}\, \Qnut^{1/2}\ ,\qquad
M=\fft{\Qw}{2\kappa_3^2\, \sqrt{Q_e\, Q_m}}\ ,\nn\\
&&R^2 = r^2 + 2 \kappa_3^2\, M\, \ell^2\ .
\label{units}
\eea
This is precisely the extremal BTZ black hole solution~\cite{btz},
{\it i.e.}  a rotating black hole solution of three-dimensional
Einstein gravity with a negative cosmological constant, described by
the Lagrangian $\kappa_3^2\, e^{-1}\, {\cal L} = R - 2 \ell^{-1}$.
Note that the extremal BTZ black hole is an example of a 
generalised Kaigorodov
metric, specialised to $D=3$ \cite{cvlupo}.  The entropy is given by
\be
S={{\omega_1}\over {4\kappa_3^2}}=\fft{\pi}{\kappa_3}\, 
\sqrt{2\ell^2\, M}\ .\label{ent3}
\ee
Since the coordinate $x$ has to be periodic with the period
$L_x=2\pi\, \ell$, the three-dimensional gravitational constant
$\kappa_3^2$, when expressed in terms of $L_x$ and either 
$\kappa_4^2$ or $\kappa_6^2$, is given by:
\bea
\kappa_3^2 &=& \fft{\kappa_6^2}{8 (Q_e\, Q_m)^{3/4}\, \Qnut^{1/2}
\, \omega_3}\nn\\
&=& \fft{L_x\, \kappa_4^2}{(Q_e\, Q_m)^{3/4}\, \Qnut^{1/2}\,
\omega_2}\\
&=& \fft{4\pi\, \kappa_4^2}{(Q_e\, Q_m)^{1/2}\, \omega_2}\ .\nn
\label{k3}\eea
In particular we see that when $\kappa_3^2$ is expressed in terms of
$\kappa_4^2$, it is independent of
$\Qnut$.  Note that the three-dimensional gravitational constant
$\kappa_3$ is related to the Newton's constant $G$ defined in
\cite{Strominger} by $\kappa_3^2 = 2G$.

Substuting $M$ and $\ell$ from (\ref{units}), and $\kappa_3$ from
(\ref{k3}), into the extremal BTZ entropy (\ref{ent3}) reproduces
precisely the entropy of the four-dimensional 4-charge black hole,
given in (\ref{d4ent}), as one would expect.  Thus the microscopic
counting in~\cite{Strominger,dublinbtz}, which reproduces precisely the
BTZ entropy (\ref{ent3}) in terms of the asymptotic two-dimensional CFT
with the $SL(2,R)_L\times SL(2,R)_R$ isometry \cite{adsc} ({\it via}
the AdS/CFT correspondence), in its turn reproduces the microscopic
entropy formula of the four-dimensional 4-charge black hole as well!
      
      Here we should like to comment on the ranges of the various
charges for which the above microscopic counting is valid.  The
discussion of the entropy of the four-dimensional 4-charge black hole
splits in two parts.  The first step, in which the near-horizon
geometry of the $D=4$ black hole is mapped to the BTZ black hole in
$D=3$, can be implemented when the gravity decoupling limit is valid.
This is discussed in detail in \cite{cvlupo}, and can be specified
roughly by (\ref{dec}). This is the condition for the field theory on
the intersecting $p$-brane configuration to decouple from gravity.
The second step makes use of Cardy's entropy formula
for two-dimensional CFT \cite{Cardy},  leading to the microscopic
state-counting formula for the entropy of the BTZ black hole: 
\be S=2\pi\sqrt{\ft16 c\, N_{\sst R}}
+ 2\pi\sqrt{\ft16 c\, N_{\sst L}}\ .  
\ee 
Here the central charge is given by $c= 3\ell/\kappa_3^2$, and the
Virasoro level numbers $N_L\equiv L_0$ and $N_R\equiv {\bar L}_0$ are
related to the BTZ mass $M$ and angular momentum $J$ by $L_0+\bar L_0
= M\, \ell$ and $L_0 -\bar L_0 = J$ \cite{Strominger}.  Cardy's
formula is valid only in the asymptotic limit where the growth of the
numbers of states is such that $N_{\sst{L}} + N_{\sst{R}}>> c$.  This
constraint implies that we must have
\be
\fft{M\,\ell}{3\ell/\kappa_3^2} =
\fft{\Qw}{6\sqrt{Q_e\, Q_m}} >>1\ .\label{conformalcon}
\ee
Thus we see that in order to have a conformal-field-theoretic
microscopic interpretation for the entropy, the momentum of the wave
must be very large.  Note, however, that if we nevertheless blindly
apply Cardy's formula for the case $1\ll N_{L,R}\le c$, {\it i.e.}
$Q_{\rm wave}\le (Q_e,Q_m)$, we still precisely reproduce the
classical results!
          
         It is important to note that the constraint
(\ref{conformalcon}) is independent of the value of the Taub-NUT
charge.  A particular case corresponds to the choice $\Qnut=1$, for
which the six-dimensional near-horizon geometry is precisely
BTZ$\times S^3$.  This is exactly the same as the near-horizon
geometry of the boosted dyonic string, which gives rise to the
3-charge Reissner-Nordstr\"om-type black hole in $D=5$.  (Its
microscopic state counting, using the CFT/AdS correspondence, was
understood for static black holes in \cite{Strominger}, and for
rotating black holes in \cite{cvla}.)

      To summarise, we have seen that from the six-dimensional point of
view the near-horizon geometries of the 3-charge $D=5$ and 4-charge
$D=4$ black holes are given by
\bea
D=5:&& {\rm BTZ}\times S^3\nn\\
D=4:&& {\rm BTZ} \times S^3/\Qnut\ .
\eea
Thus the oxidation to six dimensions of the near-horizon geometry of
the five-dimensional black hole can be viewed as a special case of the
oxidation of the four-dimensional black hole, in which $\Qnut=1$.
Since $\kappa_3$, when expressed in terms of $\kappa_4$, is
independent of $Q_{\Qnut}$, the above analysis shows that the
microscopics of the five-dimensional 3-charge black hole precisely
reproduce those of the four-dimensional 4-charge black hole, in the
case where the fourth charge comes from the reduction on the $U(1)$
fibre coordinate of the lens space $S^3/Z_{\Qnut}$.  In other words,
we have
\be
S_{\sst D=4} =\sqrt{\Qnut}\, S_{\sst D=5}\ .
\ee
Note that the implications are not only for the (asymptotic)
microscopic counting of states, but also also for the whole black hole
spectrum.  (The microscopic counting of the four-dimensional black
hole entropy was also discussed in \cite{jkm} using D-brane
techniques.  However, the counting was only valid for $\Qnut=1$, which
corresponds to, in essence, to 5-dimensional black holes.)

\subsection{Non-extremal case}

  We now turn to the consideration of non-extremal solutions,
highlighting the new features that arise here.  The non-extremal
4-charge black hole solution can be found in \cite{cvyoII}.  In terms
of our field configuration, it is given by
\bea
&&ds_4^2 = -(H_e\, H_m\, K\, U)^{-1/2}\, e^{2f}\, dt^2 +
(H_e\, H_m\, K\, U)^{1/2}\, (e^{-2f}\, d\rho^2 + \rho^2\, 
d\Omega_2^2)\ ,\nn\\
&& A_{\1 16} =\coth\mu_e\, H_e^{-1}\, dt\ ,\qquad
A_{\1 17} = Q_m\, \Omega_\2\ ,\nn\\
&& \cA^6_\1 = \coth\muw\, K^{-1}\, dt\ , \qquad
\cA^7_\1 = \Qnut\, \Omega_\2\ ,
\eea
where the functions $H_e$, $H_m$, $K$, $U$ and $f$ are
\bea
H_e= 1 + \fft{k\, \sinh^2\mu_e}{\rho}\ ,&&
H_e= 1 + \fft{k\, \sinh^2\mu_m}{\rho}\ ,\nn\\
K= 1 + \fft{k\, \sinh^2\muw}{\rho}\ ,&&
U= 1 + \fft{k\, \sinh^2\mun}{\rho}\ ,\qquad
e^{2f} = 1 - \fft{k}{\rho}\ .
\eea
The four charges are given by
\bea
Q_e=\ft12k\, \sinh2\mu_e\ ,&& 
Q_m=\ft12k\, \sinh2\mu_m\ , \nn\\
\Qw=\ft12k\, \sinh2\muw\ ,&& 
\Qnut=\ft12k\, \sinh2\mun\ .
\eea

     The Hawking temperature and entropy are of the form
\bea
T&=& (4\pi\, k\, \cosh\mu_e\, \cosh\mu_m\, \cosh\muw\, 
\cosh\mun)^{-1}\ ,\nn\\
S&=& \fft{k^2\, \omega_2}{4\kappa_4^2}\, 
\cosh\mu_e\, \cosh\mu_m\, \cosh\muw\,\cosh\mun\ .\label{d4nonextent}
\eea

\subsubsection{$D=6$ boosted dyonic string with a NUT charge}

       As in the extremal case, we may oxidise the 4-charge solution
in $D=4$ back to $D=6$, to obtain the metric
\bea
ds_6^2 &=& (H_e\, H_m)^{-1/2}\, 
\Big( -K^{-1}\, e^{2f}\, dt^2 + K\, (dx + \coth\muw\,
(K^{-1} -1)\, dt)^2\Big)\nn\\
&&+(H_e\, H_m)^{1/2}\, 
\Big(U\, (e^{-2f}\, d\rho^2 + \rho^2\, d\Omega_2^2)+
U^{-1}\, (dy + \Qnut\, B)^2\Big)\ .\label{d6nonext}
\eea
We shall be concerned with the near-extremal regime, which is defined
by taking $k$ to be small, with $\mu_e$, $\mu_m$ and $\mun$ large, so
that
\be
Q_e \sim \ft14 k \, e^{2\mu_e}\ ,\qquad
Q_m \sim \ft14 k \, e^{2\mu_m}\ ,\qquad
\Qnut \sim \ft14 k \, e^{2\mun}\ ,
\ee
are all finite and non-vanishing. 
It follows that in the near-horizon region $\rho\rightarrow 0$, 
the ``1'' in the functions $H_e$, $H_m$ and $U$ 
can be dropped, and we have 
$k\sinh^2\mu_e\sim Q_e$, $k\sinh^2\mu_m\sim Q_m$ and
$k\sinh^2\mun\sim \Qnut$. Note that we do not impose any restriction
on $\mu_{\rm wave}$. Making the coordinate transformation
\be
\rho=\fft{r^2}{4k\sinh^2\mun}\ ,
\ee
the metric (\ref{d6nonext}) becomes
\bea
ds_6^2 &=&
-\fft{r^2(r^2-4k^2\, \sinh^2\mun)}{\ell^2\, (r^2 + 4k^2\, 
\sinh^2\muw\, \sinh^2\mun)}\, dt^2 \nn\\
&&+
\fft{r^2+4k^2\, \sinh^2\muw\, \sinh^2\mun}{\ell^2}\, 
\Big(dx - \fft{4\Qw\, k\,\sinh^2\mun}{r^2 + 4k^2\, \sinh^2\muw\, 
\sinh^2\mun}\, dt\Big)^2\nn\\
&&+  \fft{\ell^2\, dr^2}{r^2 -4k^2\, \sinh^2\mun} +
\ell^2\, \Big(\ft14\coth^2\mu_4 (\fft{dy}{\Qnut} + B)^2 +
\ft14\, d\Omega_2^2\Big)\ ,\label{d6horizonnonext}
\eea
where $\ell^2 = 4k^2\, \sinh\mu_e\, \sinh\mu_m\, \sinh^2\mun$.  We
see that this six-dimensional metric is the direct sum of two
three-dimensional metrics.  In particular, the factor
\be 
d\bar s_3^2 = \ell^2\, \Big(\ft14\coth^2\mun\,  (\fft{dy}{\Qnut} + B)^2 +
\ft14\, d\Omega_2^2\Big)\label{squash3} 
\ee 
describes a {\it squashed} three-dimensional cyclic lens space, where
the squashing parameter is $\coth^2\mun$ and the 3-sphere is factored by
$Z_\Qnut$. (We are assuming that, as usual, $y$ has period $4\pi$.) In
the extremal limit with non-vanishing
$\Qnut$, which requires that $\mun\rightarrow \infty$, the squashing
parameter
$\coth\mun$ becomes 1 and the space becomes the unsquashed lens space.
In the decoupling limit, which corresponds to the near-extremal region
with $\mun>>1$, the squashing effect is very small, and the lens
space is almost round.  (If we set $\Qnut$ to zero, and hence
$\mun=0$, the space will instead be untwisted, becoming $S^2\times S^1$.) The
volume of this squashed lens space is given by
\be
V=\fft{\pi\,\omega_2\, \ell^3\, \coth\mun}{2\Qnut} =
2\pi\, \ell\,\omega_2\, k\, \sinh\mu_e\, \sinh\mu_m\ .
\ee

\subsection{Microscopic counting and the BTZ black hole}

      Dimensionally reducing the metric (\ref{d6horizonnonext}) on the
squashed lens space (\ref{squash3}), we obtain precisely the
three-dimensional BTZ black hole, given by
\be
ds_3^2 =- \fft{r^2\,(r^2-r_+^2)}{\ell^2\, R^2}\, dt^2 +
R^2 (\fft{dx}{\ell} - \fft{2\kappa_3^2\, J}{R^2}\, dt)^2 + 
\fft{\ell^2}{r^2 - r_+^2}\, dr^2
\ ,\label{btzbh}
\ee
where
\bea
&&R^2 = r^2 + \ft12 (4\kappa_3^2\, M\ell^2 -r_+^2)\ ,\nn\\
&&\ell^2 = 4k^2\, \sinh\mu_e\, \sinh\mu_m\, \sinh^2\mun
\sim 4\sqrt{Q_e\, Q_m}\, \Qnut\ ,\nn\\
&&M\,\ell^2 = \fft{k^2}{\kappa_3^2}\, \cosh2\muw\, \sinh^2\mun
\sim \fft{\Qnut}{\kappa_3^2}\, k\,\cosh 2\muw\ ,
\quad J=M\,\ell\, \tanh2\muw\ ,\nn\\
&&r_+^2 =4\kappa_3^2\, M\ell^2\, 
\sqrt{1-(\fft{J}{M\ell})^2}=4k^2\, \sinh^2\mun
\sim 4k\,\Qnut \ .\label{btzvar}
\eea
Thus the coordinate $x$ has the period of $L_x=2\pi\, \ell$.   Following
the same discussion as in the extremal case, we find that the
three-dimensional gravitational constant $\kappa_3$ is related to that
of the original four-dimensional theory by
\be
\kappa_3^2=\fft{4\pi\,\kappa_4^2}{\omega_2\, k\, \sinh\mu_e\,
\sinh\mu_m}\sim \fft{4\pi\, \kappa_4^2}{(Q_e\, Q_m)^{1/2}\, \omega_2}
\ .
\ee

      The entropy of the BTZ black hole in this case takes the form:
\be
S=\fft{\pi}{\kappa_3}\Big(\sqrt{\ell(M\,\ell + J)} +
\sqrt{\ell(M\,\ell - J)}\Big)\ .\label{nonextcount}
\ee
Substituting the variables in (\ref{btzvar}) and $\kappa_3^2$ into the
above entropy formula, we obtain
\bea
S&=& \fft{k^2\, \omega_2}{4\kappa_4^2}\, 
\sinh\mu_e\, \sinh\mu_m\, \cosh\muw\,\sinh\mun\nn\\
&\sim& \fft{\omega_2}{4\kappa_4^2}\, \sqrt{Q_e\, Q_m\, \Qnut
\, k\, \cosh^2 2\muw}\ ,\label{btznonextent}
\eea
which is precisely the entropy (\ref{d4nonextent}) for the
four-dimensional black hole in the near-extremal region where $\mu_e$,
$\mu_m$ and $\mun$ are all much greater than 1, and $k$ tends to zero,
while keeping $Q_e$, $Q_m$ and $\Qnut$ fixed.  Note that this
agreement of the entropy formulae occurs only in the near-extremal
region.

      It is instructive to study the limit where the counting of the
string states (\ref{nonextcount}) on the boundary of the BTZ black
hole is valid.  In this non-extremal case, the central charge is given
by $c=3\ell/\kappa_3^2$, and we have $L_0 + \bar L_0 = M\,\ell$ and
$L_0-\bar L_0=J$.  The state counting gives the expression
$S=2\pi\sqrt{\ft16 c\, N_{\sst R}} + 2\pi\sqrt{\ft16 c\, N_{\sst L}}$
for the entropy, valid when $N_{\sst R} + N_{\sst L} >> c$.  This
expression is in agreement with (\ref{nonextcount}).  The constraint
on the level-numbers implies that
\be
\fft{M\,\ell}{3\ell/\kappa_3^2} =
\fft{\cosh 2\muw}{12 \sinh\mu_e\, \sinh\mu_m} \sim
\fft{\Qw}{6\sqrt{Q_e\, Q_m}} >>1\ .\label{conformalcon2}
\ee
Again, we see that this constraint is independent of the NUT charge
$\Qnut$.

\section{$p$-branes from flat space-time}

      In the previous section, we made use of the fact that the fourth
charge of the 4-charge black hole in $D=4$ can be obtained from the
Hopf reduction of the $D=5$ 3-charge black hole on the fibre
coordinate of the lens space $S^3/Z_N$, described as a $U(1)$ bundle
over $S^2$.  This is a special case of general discussion that can be
given for any $N$-charge $p$-brane solution whose transverse space is
four-dimensional; by a similar Hopf reduction we can obtain an
$(N+1)$-charge $p$-brane solution in one dimension less \cite{dlpads}.
In this section, we show that if the effect of Kaluza-Klein massive
modes is neglected, $p$-branes configurations can be obtained from
flat space-time by a sequence of dimensional reductions and oxidations,
and U-duality transformations.  This provides an alternative way of
constructing BPS p-brane solitons, without needing to go through the
process of explicitly solving the supergravity equations of motion.
In other words, the non-trivial BPS soliton solutions can be obtained
by acting with symmetry transformations on the trivial flat
space-time solution.  In this context, therefore, U dualities play the
r\^ole of solution-generating symmetries.

\subsection{$D=4$ black holes from $D=5$ Minkowski space-time}

    We begin with five-dimensional Minkowski space-time, written as
\be
ds_5^2 = -dt^2 + dr^2 + r^2\, d\Omega_3^2\ ,\label{mink5}
\ee
where $d\Omega_3^2$ is a metric on the unit 3-sphere.   Exploiting the
fact that $S^3$ is a $U(1)$ bundle over $S^2$, we may write the
3-sphere metric as
\be
d\Omega_3^2 = \ft14 d\Omega_2^2 + \ft14 (dz+B)^2 \ ,\label{s3bun}
\ee
where $d\Omega_2^2$ is the metric on the unit 2-sphere, and
$dB=\Omega_\2$, the volume form on the unit 2-sphere.  The $U(1)$
fibre coordinate $z$ has period $4\pi$.  

    Adopting the standard Kaluza-Klein ansatz for the
metric, we now reduce from $D=5$ to $D=4$ on a circle:
\be
ds_5^2 = e^{-\phi_1/\sqrt3}\, ds_4^2 + e^{2\phi_1/\sqrt3}\, 
          (dz_1 + \cA_\1^1)^2\ .\label{d5d4met}
\ee
With this ansatz, the pure Einstein-Hilbert Lagrangian ${\cal L}_5= 
R\, {*\oneone}$ in $D=5$ reduces to
\be
{\cal L}_4 = R\, {*\oneone} -\ft12 {*d\phi_1}\wedge d\phi_1 - 
          \ft12 e^{\sqrt3\phi_1}\, {*\cF_\2^1}\wedge \cF_\2^1
\ .\label{d4lag}
\ee

     We may now apply the reduction (\ref{d5d4met}) to the
five-dimensional Minkowski space-time (\ref{mink5}), which is, of
course, a solution of the pure gravity equations in $D=5$.  Writing
$d\Omega_3^2$ as in (\ref{s3bun}), we take the compactification
coordinate $z_1$ to be the Hopf fibre coordinate $z$, and the
Kaluza-Klein vector $\cA_\1^1=B$.  Thus from (\ref{d5d4met}) we obtain
the four-dimensional configuration
\bea
ds_4^2 &=& \ft12 r\Big[ -dt^2 + dr^2 + \ft14 r^2\, d\Omega_2^2\Big]\ ,\nn\\
e^{-\fft2{\sqrt3}\phi_1} &=& \fft4{r^2}\ ,
\qquad \cF_\2^1 = dB = \Omega_\2\ .\label{d4sol}
\eea
This is necessarily a solution of the equations following from the
dimensionally-reduced Lagrangian (\ref{d4lag}). 

   We now make the coordinate transformation $r=2\, \rho^{1/2}$, and
define $H=\rho^{-1}$, in terms of which the four-dimensional solution
(\ref{d4sol}) becomes
\bea
ds_4^2 &=& -H^{-1/2}\, dt^2 + H^{1/2}\, (d\rho^2 + \rho^2\,
d\Omega_2^2)\ ,\nn\\
e^{-\fft2{\sqrt3}\phi_1} &=& H \ , \qquad
\cF_\2^1 =\Omega_\2 \ .\label{d4sol2}
\eea
We observe that $H$ is a harmonic function in the flat
three-dimensional ``transverse space'' with metric $d\rho^2 + \rho^2\,
d\Omega_2^2$.  

The solution (\ref{d4sol2}) is superficially like the standard
four-dimensional single-charge extremal black hole.  The only
difference is that in (\ref{d4sol2}) the harmonic function tends to
zero at infinity, while in the usual black hole solution one has
$H=1+Q/\rho$, and the harmonic function is asymptotically constant.
In fact, although the metric ({\ref{d4sol2}) has the same structure as the
usual black hole in the near-horizon ($\rho\rightarrow 0$ limit), its
asymptotic behaviour is quite different, and in fact it has no
asymptotically Minkowskian limit.  However it has been shown that, by
any of a number of somewhat different procedures, one can use U-duality
transformations to change the values of the constant terms in the
harmonic functions in black-hole or $p$-brane solutions
\cite{hyun,bps,cllpst,bb}.  The most convenient of these for our
purposes is the one introduced in \cite{cllpst}.  This is a universal
prescription, in which one diagonally dimensionally reduces a
$D$-dimensional $p$-brane on all its world-volume dimensions
(including time), thereby obtaining a Euclidean instanton solution in
$D-p-1$ dimensions.  The dimensionally-reduced Lagrangian describing
this solution has a global symmetry group that includes a number of
independent $SL(2,\R)$ factors, one associated with each harmonic
function. In fact there is an $SL(2,\R)/O(1,1)$ scalar coset
associated with each $SL(2,\R)$ factor. By making $SL(2,\R)$
transformations on a given solution, a new one with harmonic functions
that are shifted and scaled by constants can be obtained
\cite{cllpst}.  The original motivation for transforming the harmonic
functions was in fact to strip off the constant terms, so that the
black-hole or $p$-brane solution was transformed into its near-horizon
limit.  Here, our interest lies in the opposite direction, in that we
want to transform the harmonic function $H$ in (\ref{d4sol2}) from the
degenerate form $H=\rho^{-1}$ into the standard black-hole form where
there is a constant term.  

    To apply the procedure of \cite{cllpst}, we first diagonally
reduce the solution (\ref{d4sol2}) to $D=3$, with the metric ansatz
\be
ds_4^2 = e^{-\phi_2}\, ds_3^2 - e^{\phi_2}\, (dt + \cA_\1^2)^2\ .
\label{d4d3met}
\ee
Thus we obtain the three-dimensional configuration
\bea
ds_3^2 &=& d\rho^2 + \rho^2 \, d\Omega_2^2\ ,\nn\\
e^{\phi} &=& H\ ,\qquad e^{\varphi} = 1\ ,\label{d3sol}\\
\cF_\2^1 &=& \Omega\ , \qquad \cF_\2^2 = 0\ ,\qquad \cF_{\1 2}^1 = 0\ ,\nn
\eea
where we have defined the dilatonic scalars $\phi$ and $\varphi$ by
\be
\phi= -\ft{\sqrt3}{2} \phi_1 -\ft12 \phi_2\ ,\qquad
\varphi = -\ft{\sqrt3}{2} \phi_2 +\ft12 \phi_1\ .\label{dilrot}
\ee
The dimensionally-reduced Euclidean-signature theory in $D=3$ has the
Lagrangian 
\bea
{\cal L}_3 &=& R\, {*\oneone} -\ft12 {*d\phi}\wedge d\phi - \ft12
{*d\varphi}\wedge d\varphi - \ft12 e^{-2\phi}\, {*\cF_\2^1}\wedge 
\cF_\2^1 \nn\\
&&+\ft12 e^{-\phi-\sqrt3\varphi}\, {*\cF_\2^2}\wedge \cF_\2^2
+\ft12 e^{-\phi +\sqrt3\varphi}\, {*\cF_{\1 2}^1}\wedge \cF_{\1 2}^1\ ,
\label{d3lag}
\eea
where $\cF_\2^1 = d\cA_\1^1 - d\cA_{\0 2}^1\, \cA_\1^2$, $\cF_\2^2 =
d\cA_\1^2$ and $\cF_{\1 2}^1 = d\cA_{\0 2}^1$.  The unusual signs for
the kinetic terms for $\cF_\2^2$ and $\cF_{\1 2}^1$ are the
consequence of having performed the dimensional reduction on the time
direction.  The three-dimensional configuration (\ref{d3sol}) is a
solution of the equations of motion following from this Lagrangian.
In fact, we may consistently truncate the fields $\varphi$, $\cA_\1^2$
and $\cA_{\0 2}^1$ (which in any case vanish in our solution) in the
Lagrangian (\ref{d3lag}).  In the resulting Lagrangian we then dualise
$\cA_\1^1$ to an axion $\chi$, giving the purely scalar Lagrangian
\be
{\cal L}_3 = R\, {*\oneone} -\ft12 {*d\phi}\wedge d\phi + \ft12
e^{2\phi}\, {*d\chi}\wedge d\chi\ .\label{d3lag2}
\ee
The unusual sign for the kinetic term for $\chi$ is the result of
having performed a dualisation in a Euclidean-signatured theory.
The Lagrangian (\ref{d3lag2}) has an $SL(2,\R)$ global symmetry, and
in fact the scalars parameterise the coset $SL(2,\R)/O(1,1)$.  Defining
$\tau= \chi + \jm \, e^{-\phi}$, where $\jm$ satisfies $\jm^2=1$ and
$\bar\jm = -\jm$, the $SL(2,\R)$ transformations can be written as
\be
\tau \longrightarrow \fft{a\, \tau + b}{ c\, \tau + d}\ ,
\ee
where $a d -b c=1$.  

   In terms of the dualised axion field $\chi$, the form of the
3-dimensional solution will be the same as (\ref{d3sol}), except that
now we will have $\chi = \chi_0 + H^{-1}$, where $\chi_0$ is an
arbitrary constant of integration.  After performing an
$SL(2,\R)$ transformation, we therefore obtain the new primed solution
\bea
ds_3^2 &=& d\rho^2 + \rho^2 \, d\Omega_2^2\ ,\nn\\
e^{\phi'} &=& H'\equiv 2c(c\, \chi_0+ d)\Big(1 + 
\fft{c\, \chi_0 + d}{2c\, \rho}\Big) 
\ ,\qquad e^{\varphi} = 1\ ,\label{d3solprime}\\
\chi' &=& {H'}^{-1} + \fft{a\, \chi_0 + b}{ c\, \chi_0 + d}\ , 
\qquad \cF_\2^2 = 0\ ,\qquad \cF_{\1 2}^1 = 0\ .\nn
\eea
(Quantities that are inert under $SL(2,\R)$ are written without primes.)

    Dualising the axion $\chi'$ back to a potential ${\cA_\1^1}'$, and
oxidising back to $D=4$, we obtain the new metric $ds_4^2 =
-{H'}^{-1/2}\, dt^2 + {H'}^{1/2}\, (d\rho^2 + \rho^2\, d\Omega_2^2)$.
In order to put this in the standard form, where it is asymptotic to
the canonical form of the Minkowski metric, we make the constant
general coordinate transformations $t\rightarrow (2c(c\, \chi_0 +
d))^{-1/4}\, t$ and $\rho\rightarrow (2c(c\, \chi_0 + d))^{1/4}\,
\rho$.  The final solution is given by
\bea
ds_4^2 &=& -\tH^{-1/2}\, dt^2 + \tH^{1/2}\, (d\rho^2 + \rho^2\,
d\Omega_2^2)\ ,\nn\\
e^{-\fft2{\sqrt3}(\phi_1-\phi_1^0)} &=& \tH \ , \qquad
\cF_\2^1 =(c\, \chi_0 + d)^2\, \Omega_\2 \ ,\label{d4sol3}
\eea
where the new harmonic function $\tH$ and the dilaton modulus
$\phi_1^0$ are given by
\be
\tH = 1 + \fft{(c\, \chi_0+ d)^2\, e^{-\fft{\sqrt3}{2}\phi_1^0}}{\rho}
\ ,\qquad e^{\fft{\sqrt3}{2}\phi_1^0} = 2c\, (c\, \chi_0+d)\ .
\ee

    More generally, we may introduce a second modulus parameter,
namely a constant $\phi_0$ to supplement $\chi_0$ in the $(\phi,\chi)$
system.  This can be done by rescaling the coordinates in
(\ref{d4sol2}), so that $t =e^{-\phi_0}\, t'$,
$\rho=e^{\phi_0}\,\rho'$.  Now following the same steps as before, we
arrive at the solution 
\bea ds_4^2 &=& -\tH^{-1/2}\, dt^2 +
\tH^{1/2}\, (d\rho^2 + \rho^2\, d\Omega_2^2)\ ,\nn\\
e^{-\fft2{\sqrt3}(\phi_1-\phi_1^0)} &=& \tH \ , \qquad \cF_\2^1 =(c\,
\chi_0 + d)^2\, e^{-\phi_0}\, \Omega_\2 \ ,\label{d4sol4} 
\eea
where the new harmonic function $\tH$ and the dilaton modulus
$\phi_1^0$ are given by
\be
 e^{\fft{\sqrt3}{2}\phi_1^0} = 2c\, (c\, \chi_0+d)\, e^{-\phi_0}\ ,
\qquad 
\tH = 1 + \fft{(c\, \chi_0+ d)^2\,e^{-\phi_0}\,  
   e^{-\fft{\sqrt3}{2}\phi_1^0}}{\rho}\ .
\ee
The magnetic charge of this four-dimensional black-hole solution is
given by
\be
Q=\ft1{4\pi}\, \int \cF_2 = (c\, \chi_0 + d)^2\, e^{-\phi_0}\ .
\ee
The free parameters $\phi_0$ and $\chi_0$ enable us to set the dilaton
modulus $\phi_1^0$ and the magnetic charge $Q$ to any desired values.
Note that if we oxidise this four-dimensional magnetic black-hole
solution to $D=5$, we obtain the NUT solution ({\it i.e.}\ $\R\times$
Taub-NUT)
\be
ds_5^2 = -e^{-\ft1{\sqrt3}\phi_1^0}\, dt^2 + 
e^{-\ft1{\sqrt3}\phi_1^0}\, \Big( \tH\, (d\rho^2 +\rho^2\,
d\Omega_2^2) + e^{\phi_1^0}\, \tH^{-1}\, (dz_1 +Q\, \cos\theta\,
d\phi)^2
\Big)\ .
\ee
The near-horizon limit, after making the replacement $\rho=\ft14 r^2$,
is
\be
ds_5^2 = -e^{-\ft1{\sqrt3}\phi_1^0}\, dt^2 + Q\,
e^{\ft1{2\sqrt3}\phi_1^0}\, (dr^2 + \ft14 r^2\, d\wtd\Omega_3^2)\ ,
\ee
where $d\wtd\Omega_3^2 = \ft14 d\Omega_2^2 +\ft14 (dz/Q + \cos\theta\,
d\phi)^2$ is the metric on a the unit-radius lens space $S^3/Z_Q$.

     Instead of oxidising the four-dimensional black hole
(\ref{d4sol4}) directly back to $D=5$, we can first perform a
four-dimensional U-duality transformation to map the solution into one
where the charge becomes electric, and is carried by a 2-form field
strength coming from the 4-form of M-theory or the NS-NS 3-form of the
type II string.  For definiteness, let us consider the case where
after the U-duality transformation, it is $F_{\2 12}$ that carries the
electric charge.  This solution can be viewed as the vertical
dimensional reduction of a five-dimensional black hole, where the
$U(1)$ isometry on the transverse space is achieved by making a
continuous ``stack'' of black holes along the $z_7$ axis.  Viewed from
distances $r$ in the remaining transverse space that are large
compared with the period of $z_7$, one may approximate the continuous
stack by a periodic array of black holes.  As we have seen, there is
already a natural $U(1)$ isometry in the transverse space of an {\it
isotropic} five-dimensional black hole, namely the $U(1)$ of the Hopf
fibres of the foliating 3-spheres.  We can dimensionally reduce the
new solution on this fibre coordinate, and then generate a 2-charge
black hole in $D=4$, with charges carried by the field strengths
$\{F_{\2 12}, {*\cF_\2^7}\}$. Again,
the new harmonic function associated with the field strength
$\cF_\2^7$ lacks a constant term, and so we have to repeat the steps
of reducing to $D=3$ and performing an $SL(2,\R)$ transformation in
order to introduce a constant term.

A discrete Weyl rotation can then again be used in $D=4$ , to rotate
the 2-charge black hole to one that involves only the field strengths
$F_{\2 12}$ and $F_{\2 34}$.  This can be vertically oxidised to
$D=5$, followed by another dimensional reduction on the $U(1)$ Hopf
fibre coordinate of the foliating 3-spheres.  This gives a 3-charge
black hole in $D=4$, supported by the field strengths $\{F_{\2 12},
F_{\2 34}, {*\cF_\2^7} \}$.  (Again, a constant term in the new
harmonic function can be introduced by following the steps described
previously.)  This 3-charge configuration can be rotated by a Weyl
duality transformation to $\{F_{\2 12}, F_{\2 34}, F_{\2 56} \}$.
Repeating the vertical oxidation, followed by Hopf reduction, once
more, we eventually arrive at a 4-charge black hole in $D=4$.  There
are in total 630 4-charge configurations in maximal supergravity in
$D=4$, given by (\ref{d4n4}).

Note that we are {\it not} saying that there is a complete sequence of
solution-mapping symmetries that takes us from the original
5-dimensional flat space-time to the four-dimensional 4-charge black
hole.  This is because there is one step in each of the processes of
adding the second, third and fourth charges which is not implemented
purely by a symmetry transformation.  This is the step where we
vertically oxidise an $N$-charge black hole from $D=4$ to $D=5$.  If
the four-dimensional solution is literally the given black hole, with
all other four-dimensional fields vanishing, then the mathematical
process of vertical oxidation gives a uniform line of five-dimensional
black holes distributed along the $z_7$ axis.  The harmonic function
describing the black holes with therefore have a dependence of the
form $1+ Q/R$ in $D=5$, where $R$ is the radial coordinate of the
remaining 3-dimensional space transverse to the line of black holes.
In order to proceed with the next step of Hopf reduction, we need
instead to be able to consider an {\it isotropic} single black hole in
$D=5$, with harmonic function of the form $1+ Q/r^2$ where $r$ is the
radial coordinate of the full 4-dimensional space transverse to the
$t$ coordinate.  To justify this step, one first has to view the
four-dimensional black hole as an approximate solution that could be
thought of as the dimensional reduction of a periodic array of
five-dimensional black holes.  (This effectively means that one is
neglecting massive Kaluza-Klein modes in $D=4$.)  Sufficiently near to
one of the black holes in this array, it approximates to a genuinely
isotropic five-dimensional black hole, with no periodic identification
of any of the Cartesian transverse coordinates.  It is this solution,
with the transverse space then written in hyperspherical polar
coordinates, that is then used for the next step of Hopf reduction on
the $U(1)$ fibres of the foliating 3-spheres, in order to generate the
next charge in $D=4$.

The five-dimensional metric is independent of the $U(1)$ fibre
coordinate, and hence massive Kaluza-Klein modes all rigorously vanish
in the Hopf reduction.  On the other hand, a periodic array of
five-dimensional black holes does depend on the compactifying
coordinate (along the periodic axis), and so Kaluza-Klein massive
modes are non-zero after the dimensional reduction.  Thus to say that
the $N$-charge and $(N+1)$-charge four-dimensional black holes are
equivalent under this vertical-oxidation/Hopf-reduction procedure is
to ignore the discrepancies in the massive Kaluza-Klein modes in four
dimensions.  This neglect of massive modes is in the same spirit as in
the usual discussion of the U-duality symmetry group: The
Cremmer-Julia global symmetry groups in supergravities arise only when
the massive Kaluza-Klein modes are set to zero.  One can see this from
a string-theory standpoint by noting, for example, that the
56-dimensional U-duality multiplet of four-dimensional single-charge
black holes come from the vertical and diagonal dimensional reduction
of M-branes, NUTs and waves in $D=11$.  Again, the vertical reduction
steps involve the neglect of Kaluza-Klein massive modes, while the
diagonal reduction steps are on coordinates which have genuine and
exact $U(1)$ isometries.  In fact a similar philosophy to the one that
interprets all the four-dimensional black holes as coming from fewer
fundamental objects in $D=11$ allows us to interpret all the black
holes as coming from flat space in $D=5$.  In both cases, the result
follows once one ignores the massive Kaluza-Klein modes.

\subsection{D-branes and NS-NS branes from $D=11$ Minkowski space-time}

The discussion of the previous section can easily be generalised to
obtain ten-dimensional $p$-branes from $D=11$ Minkowski space-time.
Thus to begin, we consider the eleven-dimensional metric
\be
ds_{11}^2 = dx^\mu\, dx_\mu  + dr^2 + r^2\, d\Omega_3^2\ .
\ee
We again perform a dimensional reduction on the $U(1)$ fibre
coordinate in $S^3$, giving rise to D6-brane in $D=10$ type IIA.  (In
order to introduce a constant term in the harmonic function, we can
diagonally dimensionally reduce on the world-volume to $D=3$, perform
an $SL(2,\R)$ transformation, and diagonally oxidise back to $D=10$.
An alternative procedure is to apply a sequence of IIA/IIB T-duality
transformations and a type IIB S-duality transformation to map the
solution to a wave.  The harmonic function describing the wave can
then be shifted and rescaled by general coordinate transformations
\cite{hyun,bps}. Yet another possibility is to map the solution
instead to an instanton in type IIB, and perform an $SL(2,\R)$
transformation there \cite{bb}.)

   Having obtained the D6-brane in the $D=10$ type IIA theory, we can
then use the IIA/IIB T-duality that relates a D$p$-brane to a
D$(p+1)$-brane to generate all the D-branes in ten dimensions.  Using
the S-duality of the type IIB theory, we can generate the NS-NS string
and 5-brane from the D-string and D5-brane respectively.  The NS-NS
string and 5-brane are T-dual to ten-dimensional waves and NUTs
respectively.

\section*{Acknowledgements} We are grateful to Mike Duff for helpful
discussions.

\end{document}